\documentclass[10pt]{article}
\usepackage[dvips]{graphicx}
\usepackage{a4wide}

\newcommand{\be}{\begin{equation}}
\newcommand{\ee}{\end{equation}}
\newcommand{\ba}{\begin{eqnarray}}
\newcommand{\ea}{\end{eqnarray}}

\def\={\,=\,}

\def\vb0{{\bf b}_0}

\def\={\,=\,}

\begin{document}
\thispagestyle{empty}
%\begin{flushright}
%WU B 10-22 \\
%hep-ph/yymmnnn\\
%September, 12 2010\\[10mm]
%\end{flushright}

\begin{center}
{\Large\bf What did we learn about GPDs from hard exclusive electroproduction of mesons?}
\vskip 5mm

P.\ Kroll\\[1em]
{\small {\it Fachbereich Physik, Universit\"at Wuppertal, D-42097 Wuppertal,
Germany
and\\
Institut f\"ur Theoretische Physik, Universit\"at
    Regensburg, D-93040 Regensburg, Germany}}\\
\end{center}
\vskip 3mm 

\begin{abstract}
\noindent It is reported on an analysis of electroproduction of light 
mesons at small Bjorken-$x$ ($x_{\rm Bj}$) within the handbag
approach. The  partonic subprocesses, meson electroproduction off 
quarks or gluons, are calculated within the modified perturbative 
approach (m.p.a.) in which quark transverse momenta are retained. 
The soft hadronic matrix elements, generalized parton distributions 
(GPDs), are constructed by means of double distributions. The 
constraints from parton distributions and sum rules 
are taken into account. Various moments of these GPDs are compared 
to recent results from lattice gauge theories. 
\end{abstract}


%\section{The handbag approach}
It has been shown~\cite{rad96} that, at large photon  virtuality $Q^2$,
meson electroproduction factorizes in partonic subprocesses, 
electroproduction off gluons or quarks, $\gamma^* g(q)\to M g(q)$, and
in GPDs, representing soft proton matrix elements which encode the soft,
non-perturbative physics. The calculation of the subprocess amplitudes 
requires the meson's wave function, i.e.\ a second soft, non-perturbative 
function. It has also been shown that in this so-called handbag approach 
which offers a partonic description of meson electroproduction, the dominant
contribution is generated by transitions from longitudinally polarized 
virtual photons ($\gamma^*_L$) to like-wise polarized vector mesons ($V_L$). 
Other transitions are suppressed by inverse powers of the large scale, 
$Q^2$, although, as we had to learn, they are not small at experimentally 
accessible values of $Q^2$. Thus, for instance, the ratio of the longitudinal 
over the transverse cross sections for the production of $\rho^0$ or $\phi$ 
mesons is about 2 for $Q^2\simeq 4\,~{\rm GeV}^2$. Other clear signals for 
contributions from transverse photons come from asymmetries measured in 
$\pi^+$ electroproduction with a transversely polarized target \cite{hermes10} 
and from the transverse cross section for this process measured by the 
$F_\pi-2$ collaboration \cite{fpi2}.
     
In the following it is reported on an analysis~\cite{GK1,GK5} of exclusive 
meson electroproduction within the handbag factorization scheme carried 
through in the kinematical regime of large energy in the photon-proton
center of mass frame ($W\geq 4\,{\rm GeV}$), low skewness ($\xi \simeq x_{\rm
  Bj}/(2-x_{\rm Bj}) \leq 0.1$) and small momentum transfer ($-t\leq 0.6\,{\rm GeV}^2$).   
In this kinematical region the dominant helicity amplitudes for the process 
$\gamma^* p\to Vp$ where $V$ denotes a vector meson, read
\begin{eqnarray}
{\cal M}^V_{\mu +,\mu +}&=&\frac{e_0}{2}\sum_a e_a {\cal C}_V^a \left\{\langle
H\rangle^g_{V\mu} + \langle H\rangle^a_{V\mu} + \langle \widetilde{H}\rangle^g_{V\mu} 
+ \langle \widetilde{H}\rangle^a_{V\mu}\right\}\,,   \nonumber\\
{\cal M}^V_{\mu -,\mu +}&=&-\frac{e_0}{2}\frac{\sqrt{-t}}{2m}\,
\sum_a e_a {\cal C}_V^a \left\{\langle
E\rangle^g_{V\mu} + \langle E\rangle^a_{V\mu}\right\}\,.
\label{amplitudes}
\end{eqnarray}
Explicit helicity labels refer to the proton while $\mu (=0,\pm 1)$ denotes
the helicity of the photon and the vector meson. The quark flavors are denoted
by $a$ and $e_a$ is the corresponding charge. The non-zero flavor weight
factors read for the vector mesons of interest 
\begin{equation}
{\cal C}_{\rho^0}^{\,u} =-{\cal C}_{\rho^0}^{\,d} = {\cal C}_\omega^{\,u} 
={\cal C}_\omega^{\,d} =1/\sqrt{2}\,, \qquad {\cal C}_\phi^{\,s}= 1\,.
\label{flavor}
\end{equation}
The terms $\langle F\rangle$ denote convolutions of subprocess amplitudes 
and GPDs ($F=H, E, \widetilde{H}$) for the two relevant subprocesses,
$\gamma^*g\to Vg$ and $\gamma^*q\to Vq$. Explicitly the convolutions read
($i=g,a$, $x_g=0$, $x_a=-1$)
\begin{equation}
\langle F\rangle^i_{V\mu} = \sum_\lambda \int_{x_i}^1 dx\, 
{\cal H}^{Vi}_{\mu\lambda,\mu\lambda}(x,\xi,Q^2,t=0)\,F^i(x,\xi,t)\,.
\end{equation}
The helicity of the parton is labeled by $\lambda$. Note that 
$\langle \widetilde{H}\rangle^i_{V}=0$ for longitudinal photons. The GPD
$\widetilde{E}$ contributes to the transverse amplitudes only to order $\xi$
and is consequently neglected as is the contribution from $E$ to the proton
helicity non-flip amplitudes because it is proportional to $\xi^2$. 

The subprocess amplitudes ${\cal H}$ are calculated within the modified
perturbative approach \cite{botts89} in which quark transverse degrees of
freedom as well as Sudakov suppressions are taken into account. This
factorization scheme is based on work by \cite{ellis,collins}. It is assumed
in \cite{GK1,GK5} that the quarks and gluons are emitted and re-absorbed by
the proton collinearly. The quark transverse momenta, $k_\perp$, are only
taken into account in the subprocesses. This approximation is justified by the
fact that the r.m.s. $k_\perp$ of the partons inside the proton is much smaller 
(the GPDs describe the full proton) than that in the meson for which
only the rather compact valence Fock state is considered. Since the Sudakov
factor is known only in the impact parameter space \cite{botts89}, canonically 
conjugated to the transverse momentum space, the subprocess amplitudes are 
calculated in the $b$ space
\begin{equation}
{\cal H}^{Vi}_{\mu\lambda,\mu\lambda} = \int d\tau d^2b\, 
         \hat{\Psi}_{V\mu}(\tau,-{\bf b})\, 
      \hat{\cal F}^{i}_{\mu\lambda,\mu\lambda}(x,\xi,\tau, Q^2,{\bf b})\, 
         \alpha_S(\mu_R)\,{\rm exp}{[-S(\tau,{\bf b},Q^2)]}\,.
\label{mod-amp}
\end{equation} 
Their $t$ dependences are neglected for consistency since they provide
corrections of order $t/Q^2$ which are generally neglected. On the
other hand, the $t$ dependence of the GPDs is taken into account
since this $t$ is scaled by a soft parameter. The hard scattering
kernels ${\cal F}$, or their respective Fourier transform $\hat{\cal F}$, 
are calculated to leading-order of perturbative QCD including quark
transverse momenta. The explicit expressions can be found in \cite{GK1}. 
Also for the Sudakov factor $S$ in (\ref{mod-amp}) and the choice of the 
renormalization ($\mu_R$) and factorization ($\mu_F$) scales it is 
referred to these articles. In collinear approximation the amplitudes 
for transversely polarized photons and vector mesons suffer from 
infrared singularities \cite{man,teryaev}. The quark transverse momenta, 
${\bf k}_\perp$, provide an admittedly model-dependent regularization
scheme of these singularities by replacements of the  type 
\begin{equation}
    \frac1{d Q^2} \longrightarrow \frac1{d Q^2 + {\bf k}^2_\perp}
\end{equation}
in the parton propagators. Here, $d$ is a momentum fraction or a product
of two. As can be readily shown the transverse amplitudes are suppressed by
$\langle k_\perp^2\rangle^{1/2}/Q$ as compared to the longitudinal ones in
this regularization scheme.  

The (longitudinal) amplitudes for electroproduction of pions are analogous
to (\ref{amplitudes}) with the replacement of $H$ and $E$ by $\widetilde{H}$
and $\widetilde{E}$, respectively. Of course, the gluonic subprocess is not
allowed in this case. For the case of $\pi^+$ production pion exchange is to
be taken into account as well.  
 
%\section{The GPDs at small skewness}
The GPDs are constructed with the help of double distributions 
\cite{muller}. The main advantage of this construction is the 
guaranteed polynomiality of the GPDs. The double distribution is
written as a product of a zero-skewness GPD and a weight function that
generates the skewness dependence of the full GPD  ($n_g=n_{\rm sea}=2$, 
$n_{\rm val}=1$)
\begin{equation} 
f_i(\beta,\eta,t)\,=\,F_i(\beta,\xi=0,t)\,\frac{\Gamma(2n_i+2)}{2^{2n_i+1}\Gamma^2(n_i+1)}\,
\frac{[(1-|\beta|)^2-\eta^2]^{n_i}}{(1-|\beta|)^{2n_i+1}}\,. 
\end{equation}
The zero-skewness GPD is parameterized as
\begin{equation}
F_i(\beta,\xi=0,t)\,=\,{\rm e}^{b_it}\,|\beta|^{-\alpha_i't}\,h_i(\beta)\,.
\label{zero-skewness}
\end{equation}
The function $h_i$ represents the forward limit, $\xi=t=0$, of the
GPD. For $H$ and $\widetilde{H}$ the forward limits are the
phenomenologically known unpolarized and polarized PDFs, respectively.
They have to be suitably continued to negative values of $\beta$.
For the other GPDs the forward limits are parameterized as
\begin{equation}
h_i(\beta) = N_i \beta^{\alpha(0)}(1-\beta)^\gamma\,.
\end{equation}
As is well-known, at low $\beta$ the PDFs behave power-like where the powers
are given by the intercepts of appropriate Regge trajectories. It seems
plausible to generate also the $t$-dependence of the GPDs by Regge ideas and
to assume that such a Regge-like behavior holds for the other GPDs as well. 
Assuming linear Regge trajectories $\alpha_i(t)=\alpha_i(0) +\alpha_i't$ 
($i=$ g, sea, valence) and exponential $t$-dependencies of the Regge residues,
one arrives at the parameterization (\ref{zero-skewness}). For large $-t$ 
one likely needs a more complicated $t$ dependence \cite{DFJK4}.

The full GPDs are obtained by an integral over $f_i$
\begin{equation}
F^{i}(x,\xi,t)\, = \, \int_{-1}^1 d\beta\,\int_{-1+|\beta|}^{1-|\beta|}
              d\eta\, \delta(\beta+\xi\eta-x)\,f_i(\beta,\eta,t)\,. 
\end{equation}
There are other methods to generate the skewness dependence, namely
the Shuvaev transform \cite{shuvaev} and the dual parameterization
\cite{dual}. Both these methods lead to very similar results for the
GPDs at small skewness.

In Ref.\ \cite{GK1} the Regge parameters are fixed in the following
way: In agreement with the HERA data \cite{h1} on the integrated cross section
$\sigma_L$  and with the CTEQ6 PDFs \cite{cteq} the gluon (`Pomeron') 
trajectory~\footnote
{Note that the forward limit of the gluonic GPD $H$ is defined as $xg(x)$ and
  analogously for the other GPDs.} 
is taken as 
$\alpha_g=1.10+0.06\ln{(Q^2/4\,{\rm GeV}^2)}+0.15\,{\rm GeV}^{-1}t$.
The increase of its intercept with $Q^2$ is a consequence of evolution.
Since the sea quarks mix with the gluons under evolution,
$\alpha_{\rm sea}(t)=\alpha_g(t)$ is assumed. For the valence quark GPDs, $H$
and $E$, on the other hand, a standard Regge trajectory is taken - 
$\alpha_{\rm val} = 0.48+0.90\,{\rm GeV}^{-2}t$. For the other GPDs there is
no prominent Regge exchange, probably Regge cuts also play an important role.
Therefore, effective trajectories are used to parameterize the low $x$
behavior of these GPDs \cite{GK5} whose parameters are fitted to experiment 
as it is done for the slope parameters $b_i$.  

It has been checked in \cite{GK1,GK5} that the GPDs respect positivity bounds 
as well as the sum rules, i.e. their first moments are in agreement with the 
nucleon form factor data at small $-t$. The forward limit of $E$ for the 
valence quarks is chosen in agreement with the form factor analysis
performed in \cite{DFJK4}.

%\section{Comparison with experiment}
\begin{figure}[h]
\centerline{\includegraphics[width=0.40\textwidth, bb= 31 335 511 713,clip=true]
{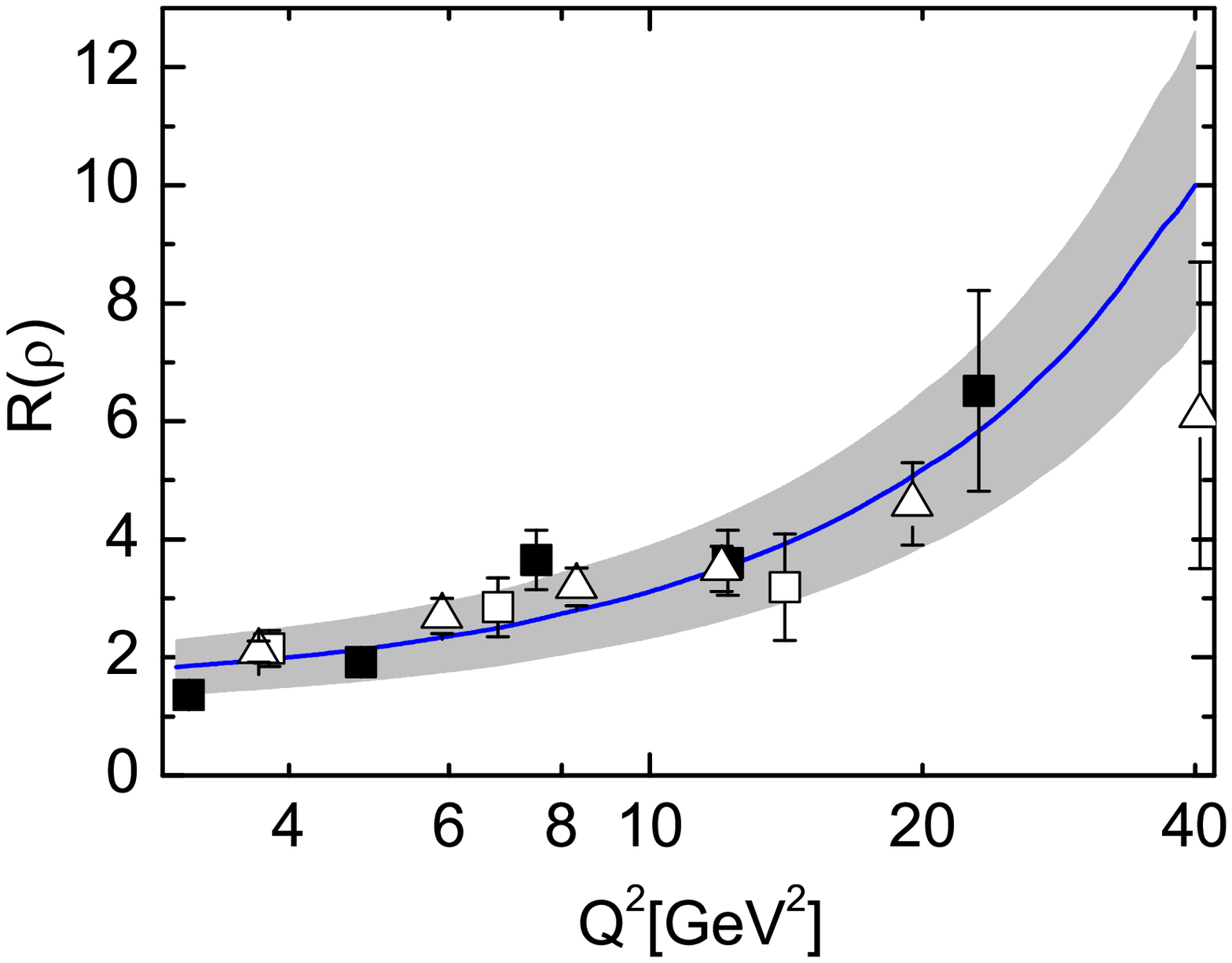}\hspace*{0.06\textwidth}
\includegraphics[width=.37\textwidth,bb=26 330 512 741,clip=true]
{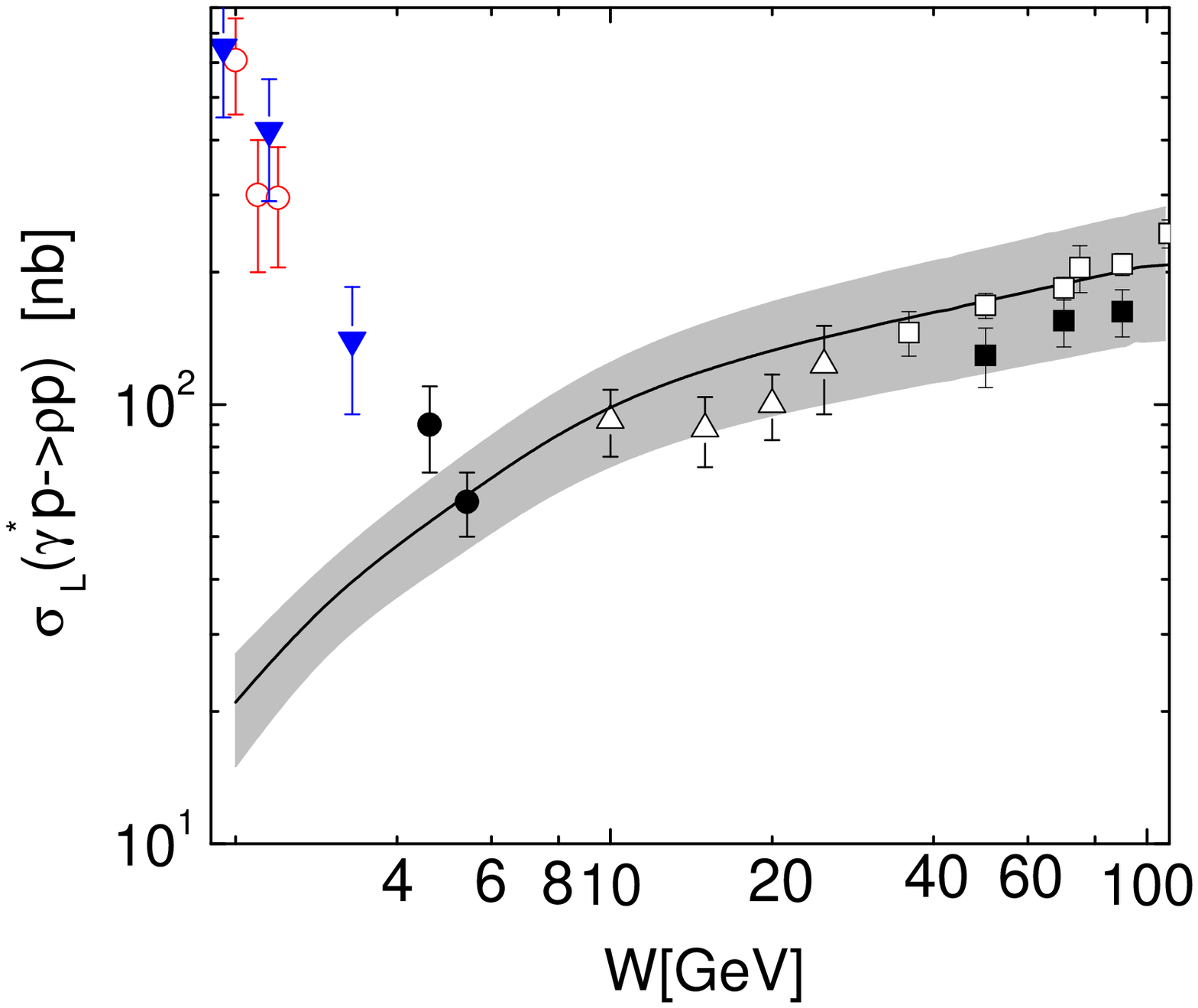}}
\caption{Left: Handbag predictions \cite{GK1} of the ratio of longitudinal and
  transverse cross sections for $\rho^0$ production versus $Q^2$ at $W=90$ GeV 
  shown as a solid line. Data taken from H1 (solid squares) and ZEUS (open
  squares and triangles). Right: Predictions of the longitudinal cross section 
  of $\rho^0$ production  versus $W$ at $Q^2=4\,{\rm GeV}^2$. The open circles 
  represent the recent CLAS data \cite{clas-rho}, the solid triangles the
  CORNELL data \cite{cornell}. The shaded bands indicate the uncertainties of 
  the theoretical analysis. For detailed references it is referred to \cite{GK1}.}
\label{Fig:1} 
\end{figure}
Given that $E$ is not much larger than $H$ and $\widetilde{H}$ much smaller
(see below) the cross sections for vector meson electroproduction are
dominated by contributions from the GPD $H$ at small skewness and small $-t$. 
An exceptions is $\rho^+$ production for which the relevant flavor
combination $E_v^u-E_v^d$ is indeed substantially larger than $H_v^u-H_v^d$
($v$ denotes valence quarks). The GPD $E$ is for instance probed by transverse
target asymmetries which are related to interference terms 
${\rm Im}[\langle E \rangle^*\langle H\rangle]$, and $\widetilde{H}$ by double
spin asymmetries like $A_{LL}$ and by electroproduction of pions.
In \cite{GK1} a detailed analysis of cross sections and spin density matrix
elements for $\rho^0$ and $\phi$ electroproduction has been performed in
the kinematical range of $W \simeq 5 - 180\, {\rm GeV}$ and $Q^2 \simeq 3 -
100\, {\rm GeV}^2$. Generally very good results have been obtained. Two example 
of the results obtained in \cite{GK1} are shown in Fig.\ \ref{Fig:1}. The
ratio of the longitudinal and transverse cross sections increases $\propto Q^2$ 
due to the power suppression of the transverse amplitude. Despite this behavior 
the ratio is not large for $Q^2$ less than about $10\,{\rm GeV}^2$ indicating
a substantial contribution from the transverse amplitude. The longitudinal
cross section for $\rho^0$ production shows a mild increase at large $W$ since
$\sigma_L\propto W^{4(\alpha_g(0)-1)}$. This signals the dominance of the gluonic 
subprocess (with a certain admixture from sea quarks). The valence quarks are 
only perceptible for $W$ smaller than about $10\,{\rm GeV}$. For HERMES 
kinematics ($W\simeq 5\,{\rm GeV}$) the gluon (plus sea) contribution still 
amounts to about $50\%$ of the cross section. Inspection of Fig.\ \ref{Fig:1} 
reveals on the other hand that the handbag approach as proposed in \cite{GK1} 
fails for low $W$: The sharp increase of $\sigma_L$ between $W=5$ and $2\,{\rm GeV}$
\cite{clas-rho,cornell} is not reproduced if the cross section is evaluated
from the above described low-skewness GPDs simply extrapolated to larger $\xi$ 
and $E$ being still neglected. The reason of this discrepancy is still
unclear. In contrast to this result an analogous extrapolation to low $W$ for
$\phi$ production leads to fair agreement with experiment \cite{clas-phi}. This 
may be regarded as a hint at a small gluonic (and sea quark) $E$.
  
\begin{figure}[hb]
\centerline{\includegraphics[width=0.43\textwidth, bb= 44 327 533 753,clip=true]
{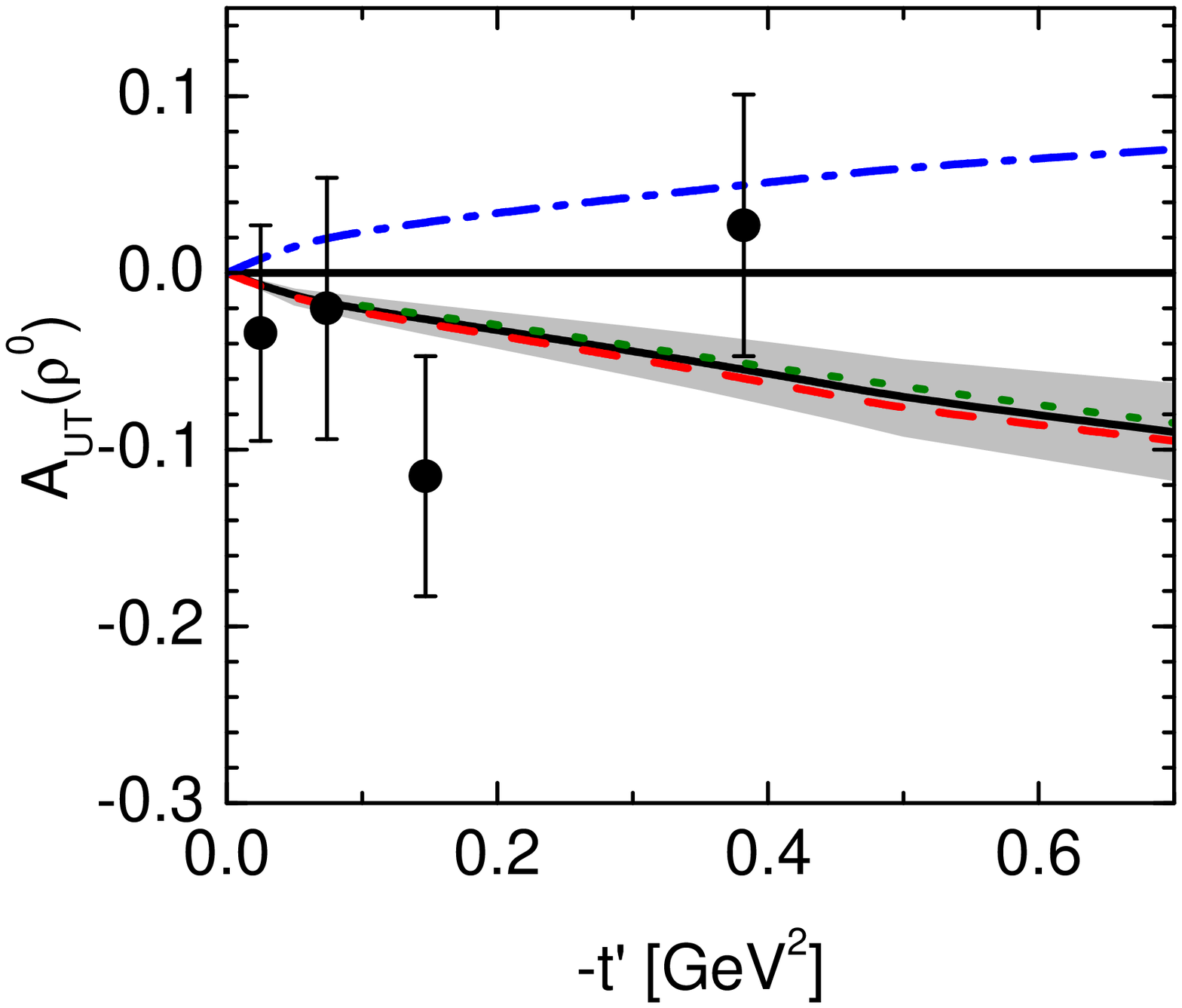}
\includegraphics[width=0.46\textwidth, bb=23 335 532 743, clip=true]
{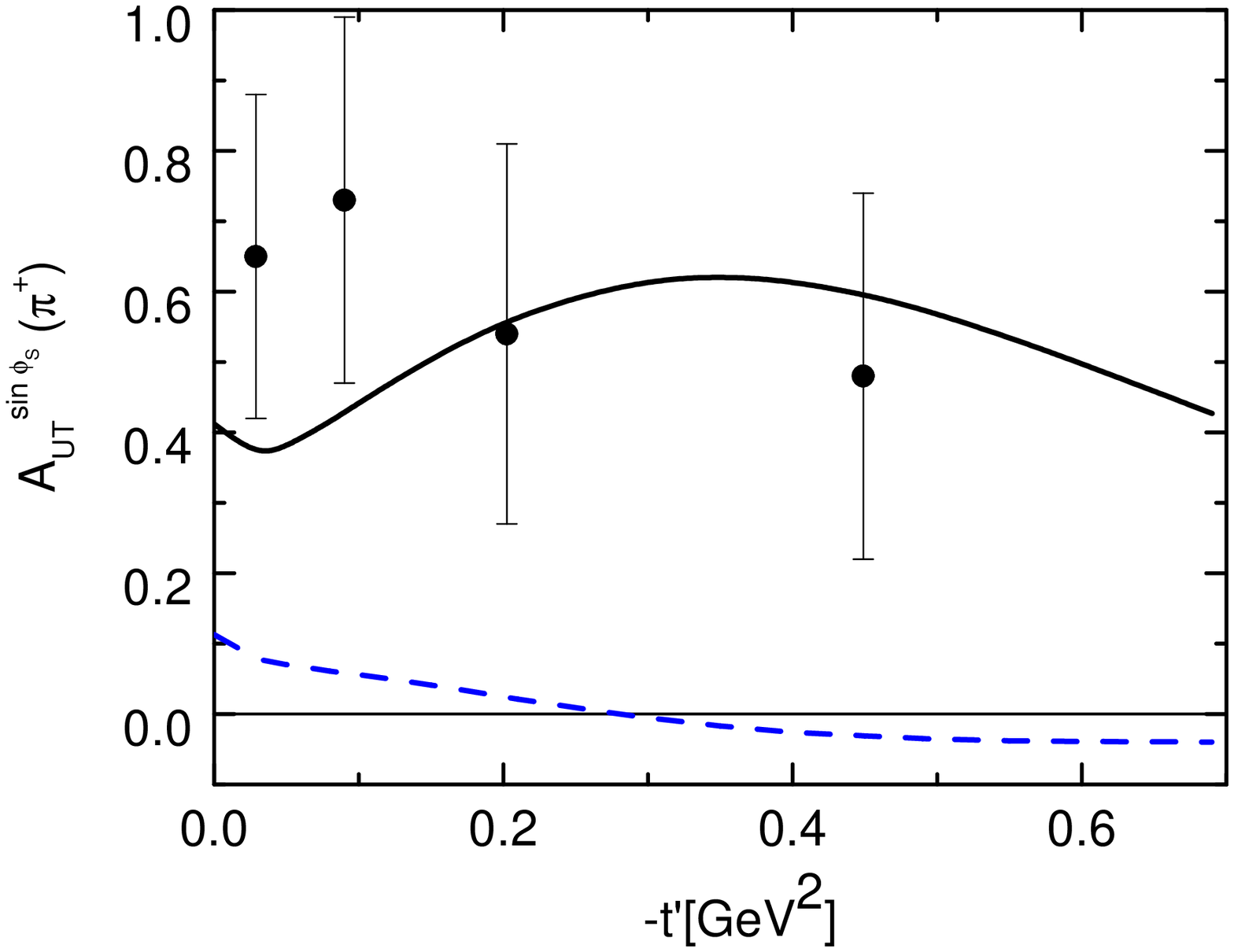}}
\caption{Left: The asymmetry $A_{UT}$ for $\rho^0$ at $Q^2=3\,{\rm GeV}^2$ 
  and $W=5\,{\rm GeV}$. Right: The $\sin{\phi_s}$ moment for $\pi^+$
  electroproduction at $Q^2=2.45\,{\rm GeV}^2$ and $W=3.99\,{\rm GeV}$. 
  The solid lines represent the predictions from the handbag approach 
  \cite{GK1,GK5,GK4}. Data taken from \cite{hermesr,hermesp}.}
\label{Fig:2} 
\end{figure}

In Fig.\ \ref{Fig:2} the transverse target asymmetry for $\rho^0$ production
is shown. The main contribution to it comes from an interference of $E$ for
valence quarks and H for gluons (plus sea). The zero-skewness GPD $E$, see
(\ref{zero-skewness}), is taken from \cite{DFJK4} (with $\alpha_{\rm val}$,
$\gamma_e^{\,u}=4.0$ and $\gamma_e^{\,d}=5.6$). Given the errors of the HERMES 
data \cite{hermesr} a reasonable fit to experiment (solid line) is obtained 
if $E^g$ and $E^{\rm sea}$ are ignored~\footnote{
Note that there are also preliminary data on this observable from COMPASS
\cite{compass}. Both the sets of data together favor negative values of
$A_{UT}$.}.
The other theoretical curves in Fig.\ \ref{Fig:2} represent results for
various variants of $E$ where $E^g$ and $E^{\rm sea}$ are estimated from
positivity bounds and a combination of Ji's sum rule and the momentum sum rule
of deep inelastic lepton-nucleon scattering \cite{DFJK4,kugler}. At present
only extreme variants seem to be excluded. It is to be emphasized that with
$E$ and $H$ at disposal one can evaluate Ji's sum rule. It turns out
\cite{GK4} that the total angular momenta of $u$ quarks and gluons are large
while those of $d$ and strange quarks are very small. The results for $u$ and
$d$ quarks are in very good agreement with lattice gauge theory \cite{haegler}.

The HERMES data on the cross section and the transverse target asymmetries for
$\pi^+$ production have been analyzed in \cite{GK5}. The relevant GPDs are
$\widetilde{H}$ and $\widetilde{E}$ as well as pion exchange. However, this is
not all. The $\sin{\phi_s}$ moment measured with a transversely polarized
target \cite{hermesp} is very large and does not seem to vanish for forward
scattering, see Fig.\ \ref{Fig:2}. Such a behavior can only be generated by
the interference of the two helicity non-flip amplitudes
\begin{equation}
A^{\sin{\phi_s}}_{UT} \sim {\rm Im}\big[{\cal M}^*_{0-,++}
{\cal M}_{0+,0+}\big]\,.
\end{equation}
As has been advocated in \cite{GK5} the amplitude ${\cal M}_{0-,++}$,
describing a $\gamma_T\to\pi$ transition, can be modeled within the handbag
approach as a twist-3 effect combining the leading-twist helicity-flip
GPDs \cite{hoodboy} with the twist-3 pion wave function \cite{braun}. Taking
into account only the most important helicity-flip GPD, namely $H_T$, one has
\begin{equation}
{\cal M}^{\rm twist-3}_{0-,++} = e_0 \langle H_T^u-H_T^d \rangle\,.
\end{equation}
The convolution is to be calculated with the subprocess amplitude 
${\cal H}_{0-,++}$ which is parametrically suppressed by $\mu_\pi/Q$ as
compared to ${\cal H}_{0+,0+}$. The parameter $\mu_\pi$ takes on a value of about
$2\,{\rm GeV}$ at a scale of $2\,{\rm GeV}$. Hence, the twist-3 effect is 
sizeable for $Q$ of the order of a few GeV. With this twist-3 effect a good
description of all HERMES data has been achieved in \cite{GK5}.

\begin{figure}
\centerline{\includegraphics[width=0.32\textwidth,bb=119 262 434 683,clip=true]
{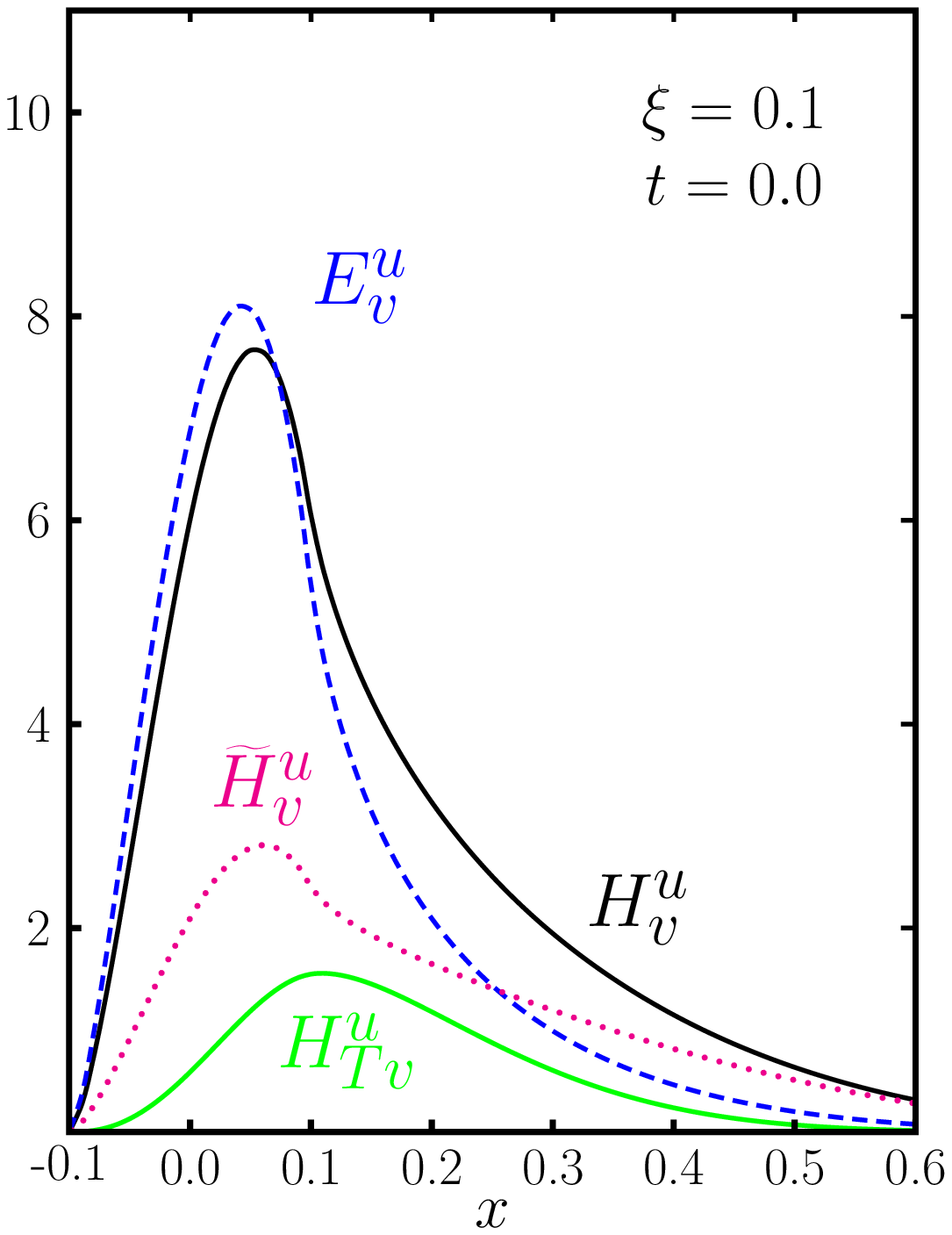}\hspace*{0.05\textwidth}
\includegraphics[width=0.323\textwidth,bb=115 263 434 684,clip=true]
{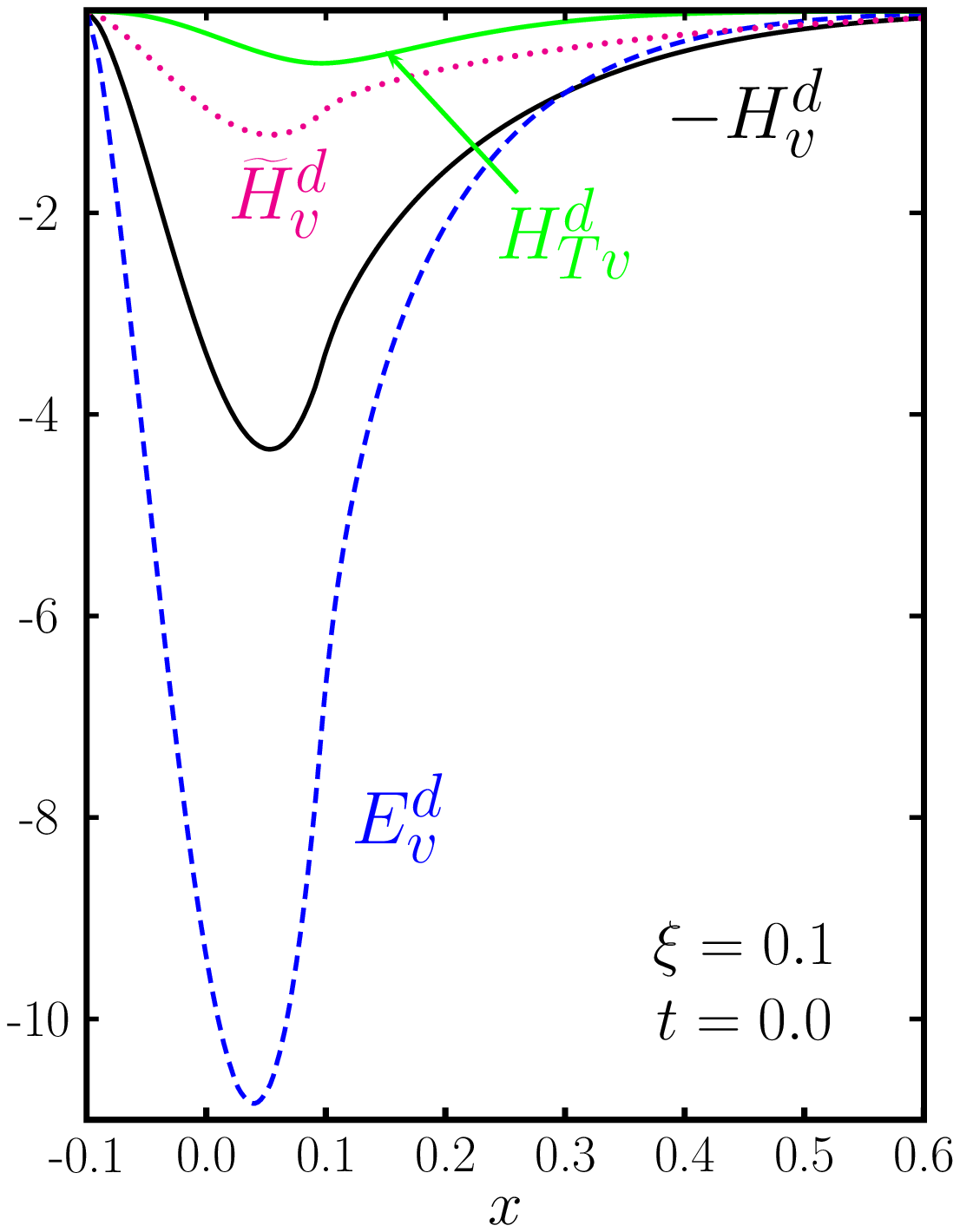}}
\caption{The valence-quark GPDs versus $x$ at $\xi=0.1$ and $t=0$. The scale
  is $4\,{\rm GeV}^2$. }
\label{Fig:3} 
\end{figure}

%\section{How do the GPDs look like?}
The GPDs extracted from meson electroproduction data in Refs.\
\cite{GK1,GK5,GK4} are valid for  
$\xi \,{\raisebox{-4pt}{$\,\stackrel{\textstyle <}{\sim}\,$}}\, 0.1\;$
and are probed by experiment for 
$x \,{\raisebox{-4pt}{$\,\stackrel{\textstyle <}{\sim}\,$}}\, 0.6\;$.
The present status of these GPDs is summarized in Tab.\ 1 and the valence
quark GPDs are displayed in Fig.\ \ref{Fig:3}.

\begin{table}[t]
\renewcommand{\arraystretch}{1.4} 
\begin{center}
\begin{tabular}{| c || c | c | c |}
\hline   
GPD & probed by &  constraints &  status \\[0.2em]
\hline
$H$ & $\rho^0, \phi$ cross sections & PDFs & known \\[0.2em]
$\widetilde{H}$  &   -   & polarized PDFs & probably small\\[0.2em]
$E$ &   $A_{UT}(\rho^0, \phi)$ & sum rule for $2^{nd}$ moment & probably
small\\[0.2em]
others &  - & - & unknown \\[0.2em]
\hline
$H$ & $\rho^0, \phi$ cross sections & PDFs, Dirac ff & known \\[0.2em]
$\widetilde{H}$  & $\pi^+$ data & pol.\ PDFs, axial ff & known \\[0.2em]
$E$ & $A_{UT}(\rho^0, \phi)$ & Pauli ff & known \\[0.2em]
$\widetilde{E}^{n.p.}$ & $\pi^+$ data & -  & uncertain \\[0.2em]
$H_T$ & $\pi^+$ data & transversity PDFs \cite{anselmino} & known\\[0.2em]
others & - & - & unknown \\[0.2em]
\hline
\end{tabular}
\end{center}
\caption{Status of small-skewness GPDs as extracted from meson
electroproduction data. The upper part is for gluons and sea quarks, the lower
part for valence quarks. Except of $H$ for gluons and sea quarks all GPDs are
probed for scales of about $4\,{\rm GeV}^2$.}
\label{tab:1}  
\renewcommand{\arraystretch}{1.0}   
\end{table}
Since their parameterizations have no nodes except at the end-points and since
they have similar $t$ dependences, their well-known lowest moments at $t=0$
fix the relative signs and strength of these GPDs.

\begin{figure}
\centerline{\includegraphics[width=0.45\textwidth,bb=158 372 590 717,clip=true]
{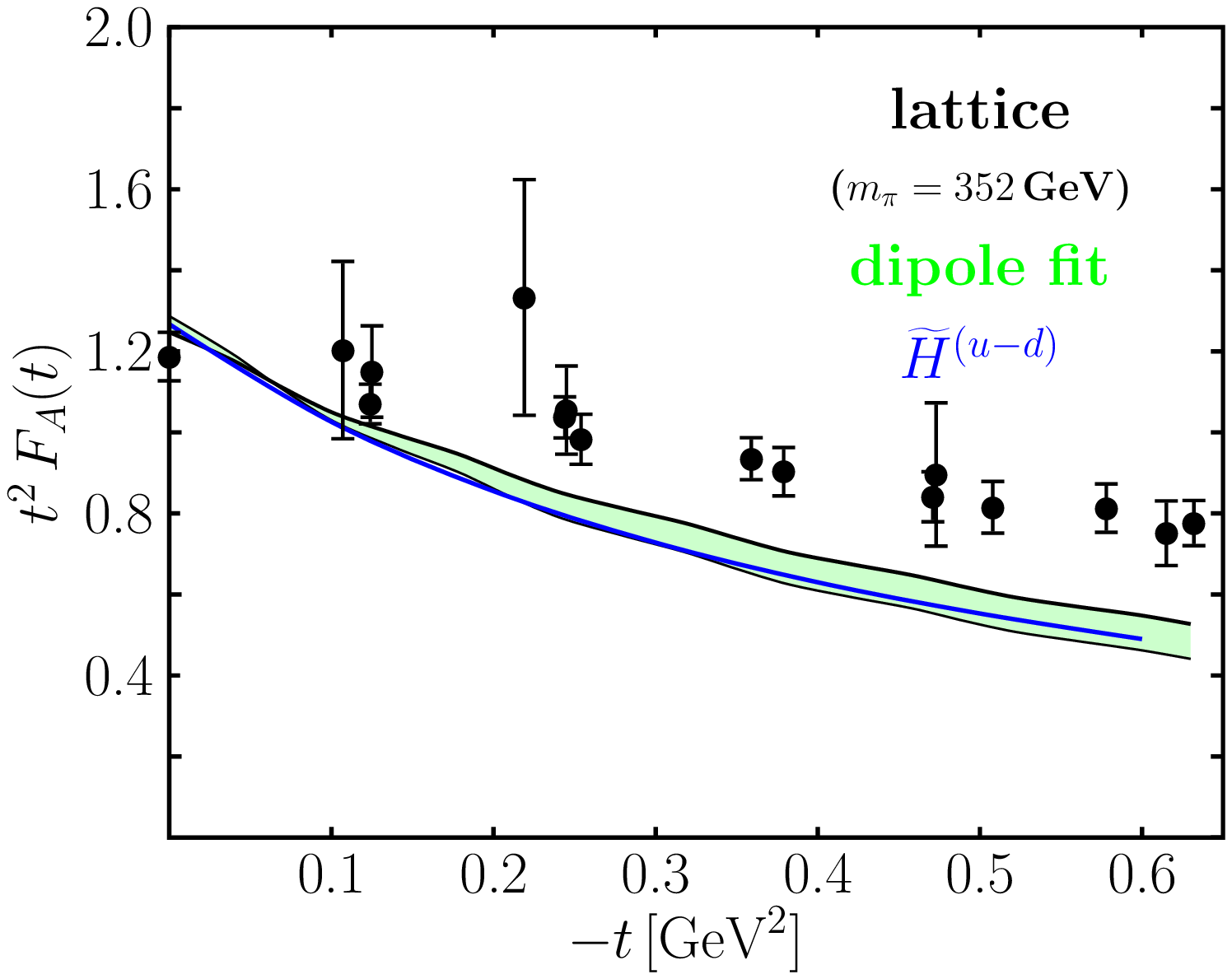}
\includegraphics[width=0.45\textwidth,bb=158 372 590 717,clip=true]
{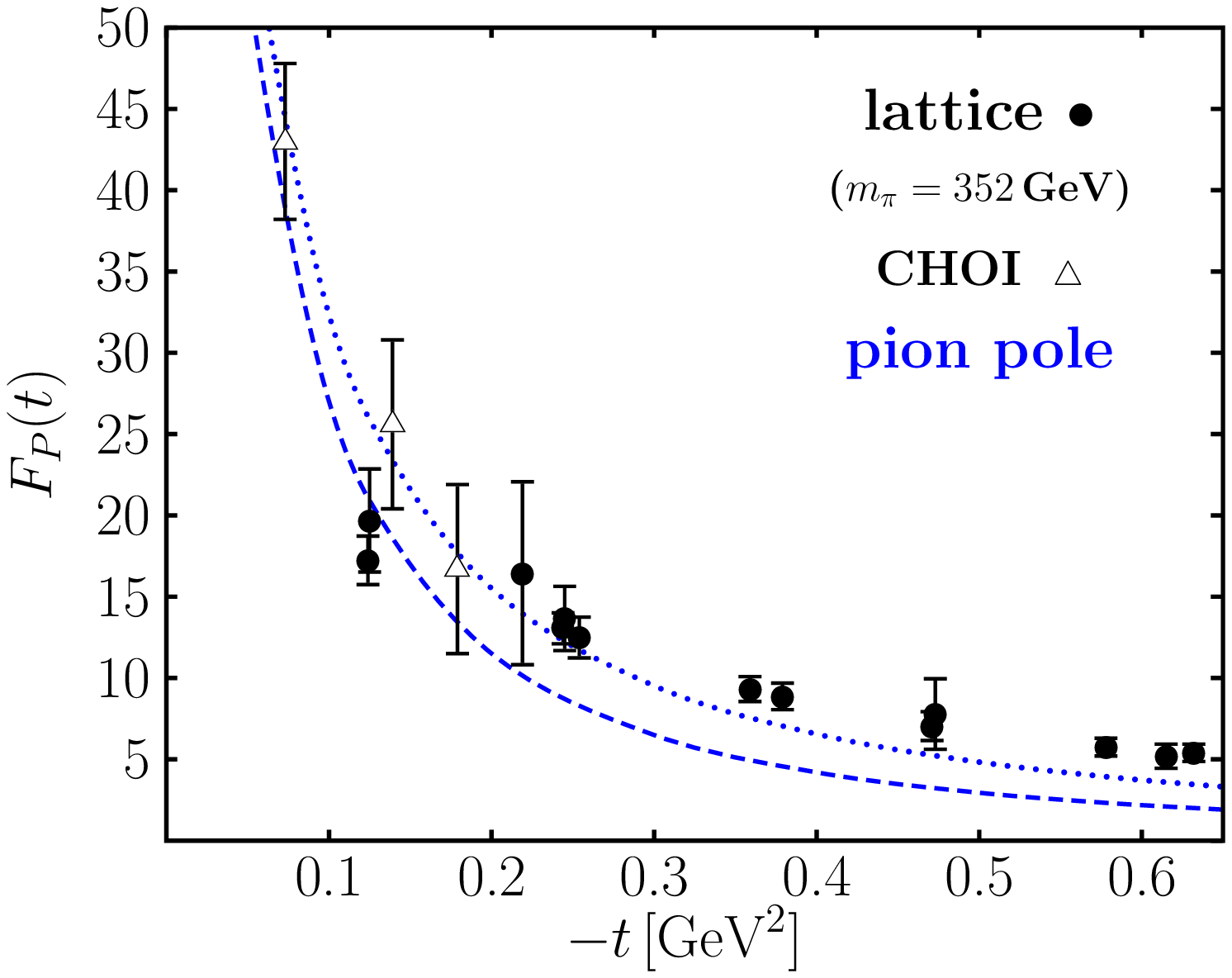}}
\caption{Left: The axial form factor of the nucleon scaled by $t^2$. The green
 band represents the dipole fit to the data \cite{kitagaki}, the solid circles 
the lattice results \cite{haegler} for $m_\pi=352\,{\rm MeV}$. The thick solid 
line is the form factor evaluated from $\widetilde{H}$.
Right: The pseudoscalar form factor of the nucleon. Experimental data from
\cite{choi}, lattice results from \cite{haegler}. The dashed (dotted) line
represents the pion-pole contribution with $\Lambda_N=0.51 (0.8)\,{\rm GeV}$.} 
\label{Fig:4}
\end{figure}

Comparison with recent lattice QCD studies~\cite{haegler, goeckeler} 
where the lowest moments of the GPDs $H, \widetilde{H}, E, \widetilde{E}$ and
$H_T$ for $u$ and $d$ quarks have been calculated,
reveals that in general there is good agreement with the relative strength of
moments and their relative $t$  dependences. At small $t$ even the absolute
values of the moments agree quite well but the $t$ dependence of the moments 
obtained from  lattice QCD are usually flatter than the form factor data and
the moments evaluated from the GPDs. An exception is the lowest moment of
$H_T$ for $u$ quarks for which a value that is about $25\%$ smaller than the
lattice result has been found in \cite{GK5}. 
In Fig.\ \ref{Fig:4} the axial-vector and the pseudoscalar form factors are
shown as examples. The form factors evaluated from the GPDs are compared to 
experimental data \cite{kitagaki,choi} and to results from  lattice QCD. For 
the pseudoscalar form factor only the pion-pole contribution
\begin{equation}
F_P^{\rm pole}(t)=4m^2 g_A\,[1-t/\Lambda_N^2]^{-1}[m_\pi^2-t]^{-1}
\end{equation}
is shown for two values of the parameter $\Lambda_N$. Since one may also
expect a flat behavior for this form factor from lattice QCD there is some 
room left for non-pole contributions from $\widetilde{E}$ at large $-t$. 

%\section{Summary}
In summary the handbag approach proposed in \cite{GK1,GK5,GK4} which consists 
of GPDs constructed from double distributions and power corrections generated 
from quark transverse momenta in the subprocess describes quite well the data 
on meson electroproduction measured by HERMES, COMPASS, FNAL and HERA over a 
wide range of kinematics. In this report the present status of this analysis 
is summarized and described what we have learned about the GPDs from it. In 
order to improve the GPDs more polarization data and data 
on $\pi^0$ electroproduction are required.

%\begin{theacknowledgments}
{\it Acknowledgments} 
This work is supported in part by the BMBF under contract 06RY258.
%\end{theacknowledgments}

\end{document}